\documentstyle[aas2pp4]{article}

\received{}
\accepted{}

\newcommand{\degr}{^\circ}

\begin{document}

\title
{A New Measurement of Cosmic Ray Composition at the Knee}

\author{K. Boothby\altaffilmark{1}, M. Chantell\altaffilmark{1}, K. D. Green\altaffilmark{1}, D. B. Kieda\altaffilmark{2}}
\author{J. Knapp\altaffilmark{1}, C . G. Larsen\altaffilmark{2},  S. P. Swordy\altaffilmark{1}}
\altaffiltext{1}{Enrico Fermi Institute, U. of Chicago, Chicago, IL 60637, USA}
\altaffiltext{2}{Department of Physics, U. of Utah, Salt Lake City, UT 84112, USA}

\begin{abstract}
The Dual Imaging Cerenkov Experiment (DICE) was designed and operated
for making elemental composition measurements of cosmic rays near
the knee of the spectrum at several PeV. Here we present the first
results using this experiment from the measurement of the average location
of the depth of shower maximum, $<X_{max}>$,
in the atmosphere as a function of particle energy.
The value of $<X_{max}>$ near the instrument threshold of $\sim 0.1$PeV is
consistent with expectations from previous direct measurements. At
higher energies there is little change in composition up to $\sim 5$PeV.
Above this energy $<X_{max}>$ is deeper than expected for a constant elemental
composition implying the overall elemental composition is becoming lighter
above the knee region. These results disagree with the idea that cosmic rays
should become on average heavier above the knee. Instead they suggest a transition
to a qualitatively different population of particles above 5 PeV.
\end{abstract}

\keywords{cosmic rays, supernova remnants, instrumentation:detectors,
ISM:abundances}

\section{Introduction}
Observations of cosmic rays have presented several puzzles which have
been resolved by detailed measurements of composition, both elemental
and isotopic. The present picture of the production of the bulk of
cosmic rays by diffusive shock acceleration in supernova remnants is
attractive since it appears to be the only energy source powerful enough to
sustain this particle population in our galaxy. There is also a strong
theoretical link between the observed energy spectra of the sources
and the predictions of strong shock acceleration,
(\cite{Axf77}, \cite{Bell78}, \cite{Bland78}, \cite{Krym77}).
However, there is still no direct evidence that the bulk of cosmic rays
are produced by this mechanism. Searches for the TeV gamma emission
which should be produced by the high energy nuclei at these sites are in progress,
(see e.g. \cite{Lessard95}).
The supernova remnant shock acceleration
in its most straightforward form produces a maximum
energy for particles accelerated by this mechanism to be close to
Z$\times$0.1PeV(\cite{Lag83}). This limit is determined by the average acceleration rate and the
mean lifetime of the strong shock in a typical supernova event.
The 'knee' in the cosmic ray flux, where the energy spectrum becomes steeper around several PeV,
could well be related to this theoretical limit.
Some workers (see e.g. \cite{Ax94}) have suggested that cosmic rays above the knee
are produced by the secondary acceleration of high energy particles from
supernova by weaker shocks in the Galaxy.
This would result in an increase in the average mass of cosmic rays through the knee.
If the cosmic rays from shock acceleration in Galactic supernovae are more strictly limited to the
knee energy then a second population of particles must be present to produce the fluxes
of cosmic rays observed at higher energies, possibly of extragalactic origin as
suggested by \cite{Protheroe92}. 
As with other properties of cosmic rays a measurement of composition
in this energy range can be used to detect these changes
and provide the information needed to investigate the origins of
cosmic rays above the knee.

While direct detection above the atmosphere
remains the most desirable method for composition determination, the
fluxes of particles in this energy range are so small that a space instrument large
enough to collect a good statistical sample up to 10PeV has not yet been
flown. Indirect detection of cosmic rays, through measurements of
air showers produced in the atmosphere, can easily supply sufficient
collecting power but these methods generally have substantially reduced
mass and energy resolution. The properties of the originating cosmic ray nucleus
becomes clouded by the huge number of interactions in the
air shower. The reconstruction of the mass of the incoming nucleus
from measurements of the shower distributions at ground level is also
subject to systematic errors resulting from imperfections of the 
Monte Carlo modelling of air showers.
DICE is a ground based air shower experiment which is designed to have
as little reliance as possible on the details of the Monte Carlo simulations
and to have the capability of comparison of results with direct measurements at
0.1PeV to provide an assessment of overall systematic errors.

\section{Experiment Description}
The two DICE telescopes  (see \cite{Booth95}, \cite{Booth97}) 
are located at the CASA-MIA site in Dugway, Utah (described in \cite{Bor94}).
They each consist of a 2m diameter f/1.16 spherical mirror with a focal plane consisting of
256 close packed 40mm hexagonal photomultipliers (PMTs) to provide $\sim 1\degr$ pixels with
a field of view of $16\degr \times 13.5\degr$ centered about the vertical.
The two telescopes are on fixed mounts separated by 100m located near the center of the CASA site.

Cosmic ray events within the field of view produce a focal plane image at the photomultiplier
cluster which is an intensity pattern of \v{C}erenkov light coming from
the air shower. When the direction of the air shower and the distance of the shower core
from the telescopes
are known, simple geometry can be used to reconstruct the amount of light received from each
altitude of the shower. The amount of \v{C}erenkov light produced at each altitude
is strongly correlated with
the number of electrons at that atmospheric depth. This is used to estimate the electron size as a function
of depth in the atmosphere from which $X_{max}$ can be determined. This procedure
is essentially geometrical and is independent of Monte Carlo simulations except for calculations
which determine the angular distribution of \v{C}erenkov light around the shower axis.
 
Here we report on the analysis of shower maxima from the images
in the DICE clusters using the independent shower geometry established
by the CASA air shower array. 
The accuracy of the CASA geometry for each shower in the knee region is $< 5$m for shower core location
on the ground
and $< 0.6\degr$ for shower trajectory direction (\cite{Bor94}). The absolute alignment of
the DICE clusters relative to the vertical
has been determined to an accuracy of $\sim 0.1\degr$ by detection of transits of stars across the DICE
apertures. 

The events which trigger both DICE telescopes and the CASA array have data which are recorded
separately and later matched by comparing the times of each event. The trigger rates are
typically 0.1Hz and the time information is accurate to $\sim 1 $ms, the fraction of
mismatched events in the raw data is $<$1\%. During analysis these mismatches are further
suppressed by (i) the requirement that shower images in DICE lie in the direction of
the CASA event, (ii) the reconstructed shower maxima from the two DICE
telescopes using the CASA geometry are in agreement. These analysis cuts further
suppress the mismatches by a factor of $\sim$1000, resulting in an overall background
level of $\sim 10^{-5}$.

\section{Data Collection and Analysis}
DICE has been operational at the Dugway site since summer 1994, this analysis comes from
the period July 1994 - December 1995. 
Data is collected during clear, moonless nights. The events analyzed here come from 914 hours of
operation, corresponding to $\sim 7$\% livetime during this period. The effective geometrical acceptance
aperture of DICE with the data cuts used 
 is $\sim$2000 m$^2$sr making the overall collecting power at high energies near
80,000 m$^2$ sr days. This allows a statistically valuable measurement of composition
up to energies of $\sim$10 PeV.
The trigger threshold and logic of the DICE PMT clusters produces some trigger
inefficiency for showers below
$\sim$800TeV, this introduces a small bias in this region which is discussed later.   

For each collected event the \v{C}erenkov images are fit by translating the PMT amplitudes
into a sequence of \v{C}erenkov luminosities and hence shower electron  sizes versus atmospheric
depth (g/cm$^2$) by using the shower geometry from CASA and the distribution of
\v{C}erenkov light about the shower axis predicted from
the CORSIKA air shower Monte Carlo (\cite{Knapp96}). These shower development curves  are
fit to a longitudinal development function from which the shower maximum, size and width
are determined. Independent fits are made for each of the two DICE clusters so that a consistency
check can made between the two sets of shower fit parameters. For the shower maxima the difference 
distribution between the fits from the two sites has an rms width of $\sim$65g/cm$^2$, after
correcting for correlations associated with the uncertainties in the geometry the dual site
rms resolution for a single event is $\sim$45g/cm$^2$. Showers which have cores located equidistant
from each cluster produce amplitudes of \v{C}erenkov light which should be similar in each 
cluster. The difference distribution of these events shows an rms width of $\sim$26\% implying a
dual site resolution of $\sim$13\% for the \v{C}erenkov size.  
The energy of each event is estimated from the mean of the independent values derived from 
the light intensity at each cluster.
The amount of \v{C}erenkov light collected at each cluster is corrected by a factor of
$B\times exp(r/r_0)$ where B is a normalization factor, $r_0$ is a scale length,
and r is the distance from the cluster to the core. The values of $r_0$ and $B$ are determined
from CORSIKA. These do not have a large dependence on the mass of the primary nucleus for the core
distances of 100m$<r<$225m used in this analysis. In this analysis the value
$r_0$=82.5m is used and the value of $B$ is taken to have a single value for all events. This
introduces a mass dependent error in the energy estimate which is $\sim$10\%. The largest uncertainty
in the energy estimate is in the conversion from PMT pulse height to \v{C}erenkov photon density
since a number of factors including mirror reflectivity and PMT quantum efficiency 
need to be accurately known. This overall systematic error in the energy scale is 
estimated to be $\pm20$\%.

The following data cuts are made:
(i) The core of the shower lies at a distance 100m$<r<$225m from both DICE clusters.
(ii)The fits of the longitudinal developments in each DICE telescope have reduced $\chi^2<3$.
(iii) The $X_{max}$ from the two sites agree within 75g/cm$^2$.
(iv) The shower direction is within $5\degr$ of the vertical.
The measurements of $X_{max}$ and energy from the two clusters are averaged to give overall
values for each shower. Values of $X_{max}$ are binned versus energy using the boundaries
shown in Table \ref{tbl-1}. To gain some confidence for the accuracy of the energy scale and the overall
detection efficiency estimates an energy spectrum of cosmic rays has been derived from the
data by calculating absolute fluxes.
The dominant efficiency correction for this work is the trigger scheme for each photomultiplier
cluster. A simulation which includes the trigger logic and air shower characteristics derived from
samples of air showers produced by CORSIKA is used to derive the various efficiencies and
the acceptance geometry of DICE. The exposure livetime 
can be directly derived from the data since at each minute of operation the cluster electronics
records the livetime for that period. The cluster deadtime is small, typically $<$1\%.
A combination of the efficiencies, geometry and livetime is used to derive the fluxes of particles
by a matrix deconvolution technique using the monte carlo.

\section{Results}
The absolute fluxes of cosmic rays obtained for these data are shown in Figure \ref{fig1}, note the fluxes are
multiplied by an overall factor of E$^{2.75}$. Also shown for comparison
are results from some previous air shower experiments which use electron sizes at ground level to
estimate the energy. The DICE data energy spectrum has been
calculated assuming a mixed composition model of the type discussed by \cite{Swo95}.
This model provides a composition at 1PeV of $\sim$40\% H, $\sim$30\% He,
and $\sim$30\% heavier nuclei and at 10PeV $\sim$20\% H, $\sim$20\% He and 
$\sim$60\% heavier nuclei.
Also shown are energy spectra reconstructed under the assumption that either
(a) the particles are all protons or (b) particles are all iron nuclei. As can be seen these do
not change the fluxes or slopes significantly. There is a distinct change in slope near $\sim$4PeV
which is consistent with the existence of the `knee' in this region. 
Since the model used to reconstruct the energy spectrum is heavier at high energies than
suggested by the values of $<X_{max}>$ measured by DICE, this could introduce a systematic bias which
might enhance the derived fluxes above $\sim$5PeV. This enhancement is at maximum 10\%,
less than the statistical precision of the fluxes in this region.
The possible effect
of a miscalibration
of the energy scale is shown by the length of the vertical bar to the left of the picture. This
represents the overall systematic uncertainty in flux from the DICE measurements corresponding
to an energy scale shift of $\pm 20$\%. The size of the trigger inefficiency correction becomes
large below $\sim$400TeV making fluxes derived from events in this region more uncertain than this
systematic limit. The overall DICE trigger efficiency is a combination of the cluster trigger efficiency and
the total light intensity produced at the cluster. DICE is therefore capable of triggering with high
efficiency on bright events with energies $>$2PeV out to the maximum extent of the data cuts at $r=225$m.
The light from lower energy particles showers 
is not intense enough to always provide triggers at this distance, the
trigger simulation is used to calculate the amount of data lost by this effect. At 500TeV the
overall detection efficiency is $\sim$50\%.

The variation of $<X_{max}>$ with energy is given in Table \ref{tbl-1}. The third
column shows the raw values and the next column shows the values of the correction, $\delta$, for trigger
and geometry bias.
The trigger bias arises from the fact that near threshold protons are more likely to trigger
DICE than iron nuclei because they produce more \v{C}erenkov light. The size of this correction
is derived from the trigger simulation which incorporates information from CORSIKA on the variation of the
intensity of light with incoming particle energy and mass. The geometry bias
of this correction is associated with the finite pixel resolution of DICE.
At high altitudes the shower structure becomes
difficult to determine because the shower development occurs on angular scales which are comparable to
the DICE pixel resolution.
The combination of these effects produces an overall correction which is relatively large for the first
energy bin but which becomes small at higher energies. These data are plotted in Figure \ref{fig2}, the
vertical errors are statistical, the horizontal correspond to the systematic error in energy estimate
discussed above.
Also shown 
is the results of direct measurements at 0.1PeV converted by the CORSIKA Monte Carlo to a value which
corresponds to the expected mean shower maximum for this composition. The small dotted lines show how the
mean shower maximum should behave for pure protons or iron nuclei based on CORSIKA.
The DICE results lie
between these extremes and agree well with the extrapolation from 
direct measurements at the lower energies, shown as the large dashed line.
There is little apparent change in mass composition through the energy range of most of these data.
The indications are that the composition becomes lighter above $\sim$5PeV although the statistics
are limited. The comparison of the measured distribution widths of $X_{max}$  
with the widths of simulations for pure protons and pure iron nuclei including the DICE detector response
are given in Table \ref{tbl-2}. These are consistent with the movement
of the mean shower maximum with energy.

The first results from DICE show no evidence for an increase in the average mass of cosmic rays in the
knee region.
These lower energy results of DICE are consistent with recent measurements which use
the {\it intensity distribution} of \v{C}erenkov 
light at
ground level to show no change in composition up to 1PeV (\cite{Plag95}).
An increase in mass has been expected somewhere
near the knee because any process which involves a steepening of magnetic rigidity (momentum/charge)
spectra of the particles,
either in the sources or elsewhere,
naturally leads to an overall increase in the mass of a population of
particles measured by energy/particle. 
It is difficult to escape the conclusion from Figure \ref{fig2} and Table \ref{tbl-2}
that a simple rigidity steepening of this type is probably not responsible for the `knee' of the
cosmic ray spectrum. 
Although these measurements indicate a transition to a more proton-rich source of particles,
in the present results this is only a $\sim 3 \sigma$ effect which should be verified by further measurements.
If confirmed, this transition suggests we are reaching an energy scale where conventional
supernova shock acceleration and propagation through the galaxy is no longer the dominant source.
We may be beginning to see a population of much older particles confined in the halo of our galaxy
or possibly of extragalactic origin. 

\section{Acknowledgements}
The authors gratefully acknowledge the assistance of the HiRes and the 
CASA-MIA collaborations and R. Lawrence. We thank M. Cassidy for technical support,
and the Commander and staff of the U.S. Army Dugway Proving Grounds
for assistance. JK acknowledges support from the Alexander von Humboldt Foundation.
This work is supported through the Block Fund
of the University of Chicago and institutional support and 
by the National Science Foundation Grant \# PHY 9514193 at the University of Utah.

\onecolumn
\clearpage
\begin{deluxetable}{ccccccc}
\tablecaption{DICE measurements of $<X_{max}>$ \label{tbl-1}}
\tablewidth{0pt}
\tablehead{
\colhead{Energy bin} & \colhead{Events}   & \colhead{$<X_{max}>$ raw }   & \colhead{$\delta$\tablenotemark{a} } & \colhead{Median Energy\tablenotemark{b}} &
\colhead{$<X_{max}>$} \\
\colhead{(PeV)} & & \colhead{(g/cm$^2$)} & \colhead{(g/cm$^2$)} & \colhead{(PeV)} & \colhead{ (g/cm$^2$) }  
} 
\startdata
0.2 - 0.4 & 6132 & 474  &-12 &0.25&462$\pm$2   \nl
0.4 - 0.7 & 3708 & 479  &-1  &0.5 &478$\pm$2   \nl
0.7 - 1.5 & 2279 & 501  &+2   &1.0 &503$\pm$2   \nl
1.5 - 3.0 & 740  & 534  &+3   &2.0 &537$\pm$3   \nl
3.0 - 5.0 & 212  & 549  &+3   &4.0 &552$\pm$4   \nl
5.0 - 8.0 & 70   & 587  &+3   &6.0 &590$\pm$10   \nl
8.0 - 15.0 & 37  & 628  &+3   &10.0 &631$\pm$14  \nl 
\enddata
 
\tablenotetext{a}{This trigger correction is discussed in the text}
\tablenotetext{b}{The energy scale has an estimated systematic error
of $\pm$20\%}
 
\end{deluxetable}

\begin{deluxetable}{cccc}
\tablecaption{DICE measurements of the widths of $X_{max}$ compared with simulations for protons and iron nuclei \label{tbl-2}}
\tablewidth{0pt}
\tablehead{
\colhead{Median Energy\tablenotemark{a}} & \colhead{DICE $\sigma(X_{max})$} & \colhead{Sim. $\sigma(protons)$} &
\colhead{Sim. $\sigma(iron)$} \\
\colhead{(PeV)} & \colhead{(g/cm$^2$)} & \colhead{(g/cm$^2$)} & \colhead{ (g/cm$^2$) }  
} 
\startdata
1.0 & 73$\pm$1 & 87 & 66  \nl
2.0 & 66$\pm$2 & 83 & 49  \nl
4.0 & 58$\pm$4 & 82 & 41  \nl
6.0 & 82$\pm$12 & 81 & 40  \nl
10.0 & 85$\pm$20 & 79 & 40  \nl
\enddata
 
\tablenotetext{a}{Only energies where the trigger bias is small are included}
 
\end{deluxetable}



\clearpage

\plotone{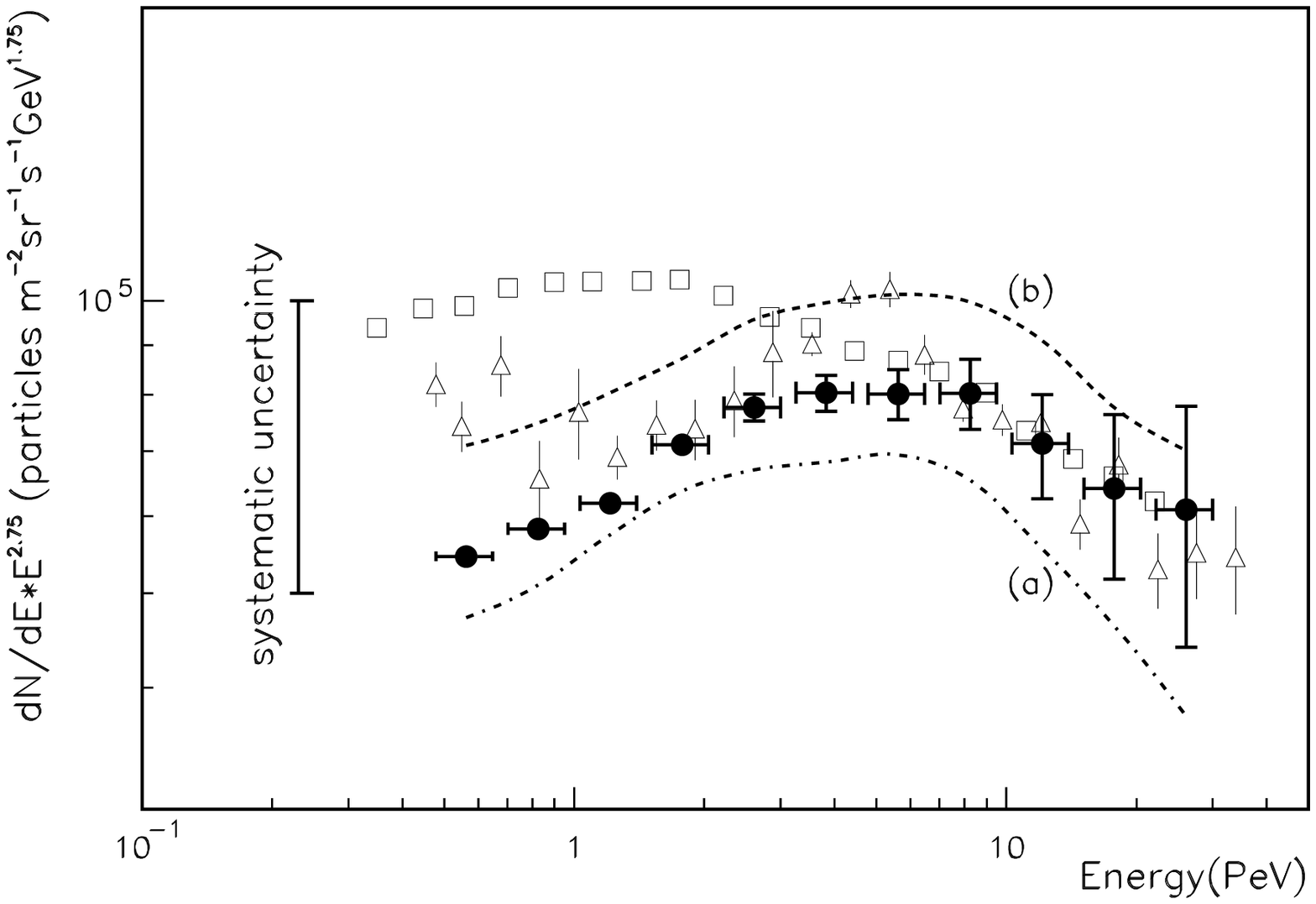}

\figcaption[specfig.eps]{The absolute fluxes of particles measured by DICE as filled circles compared
with previous measurements (triangles -\cite{Nag84}, squares -\cite{Amen96}).
Note the fluxes are multiplied by a factor of (Energy)$^{2.75}$.
These fluxes were derived
assuming the mixed elemental composition from \cite{Swo95}. Curve (a) shows the fluxes produced by assuming
an all-proton composition, curve (b) shows fluxes assuming an all-iron composition.
The vertical bar on the left represents the systematic variation of the flux levels because of
uncertainty in the overall energy scale. The DICE fluxes have power law indices of -2.55$\pm 0.01$
below 4PeV and -2.92$\pm 0.1$ above 4PeV \label{fig1}}

\clearpage

\plotone{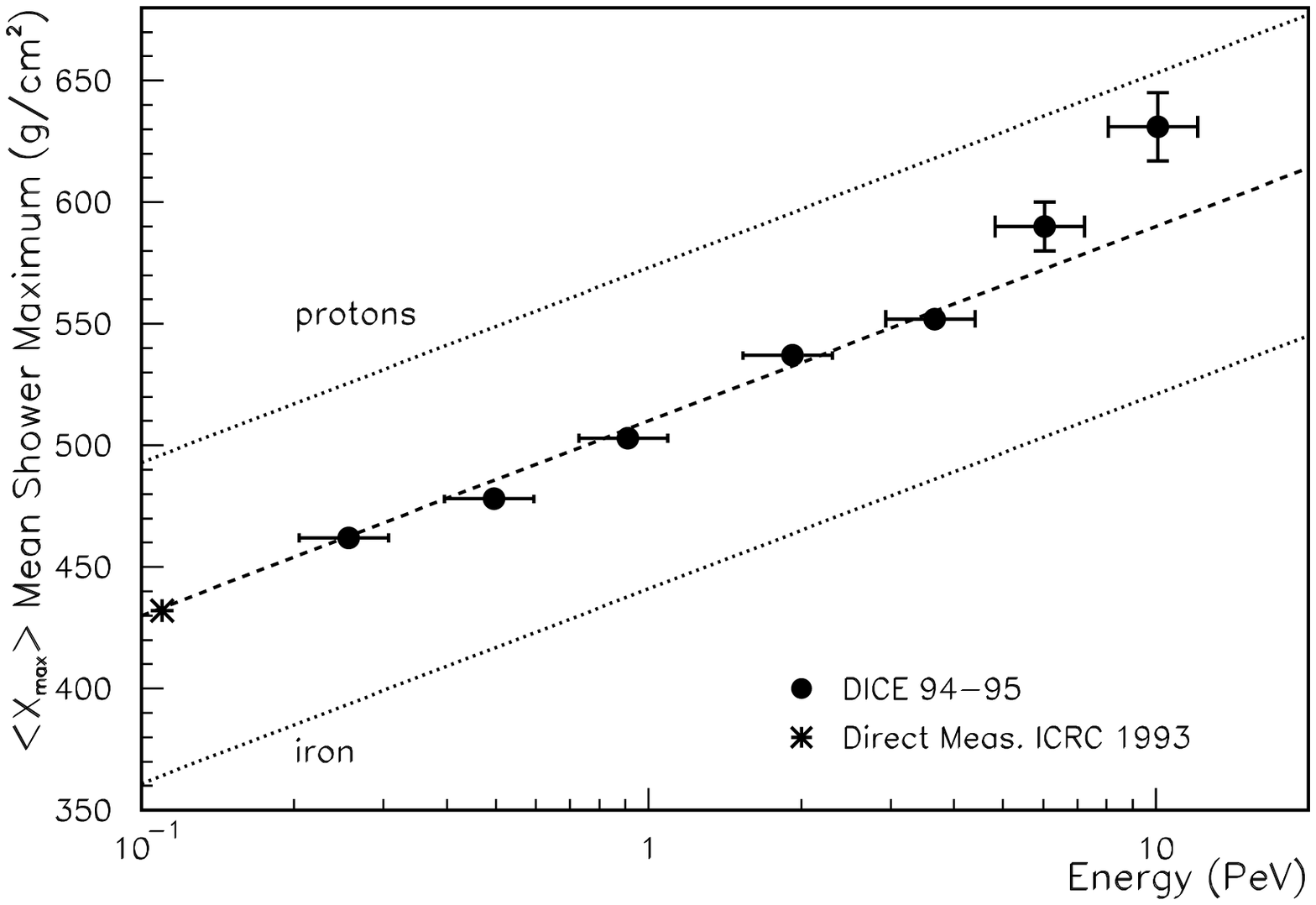}

\figcaption[xmaxfig.eps]{The variation of the mean depth of shower maximum ($<X_{max}>$) with particle energy
measured by DICE. The star is a conversion of an assembly of direct measurements (\cite{Swo94})
to the expected $<X_{max}>$ using the CORSIKA Monte Carlo - the long dashed line shows how
$<X_{max}>$ scales with energy for a constant elemental composition. The upper
dotted line is the $<X_{max}>$ expected for all-proton composition and the lower dotted
line is the value expected for all-iron composition. \label{fig2}}

\end{document}